\documentclass[runningheads]{llncs}

\usepackage{graphicx}
\usepackage{times}  
\usepackage{helvet}  
\usepackage{courier}  
\usepackage{url}  
\usepackage{amsfonts}
\usepackage{amssymb}
\usepackage{amsmath}
\usepackage{hyperref}
\usepackage{algorithm}
\usepackage{algorithmic}
\usepackage{multirow}
\usepackage{color}
\usepackage{bbding}

\def\ba{\begin{array}}
\def\ea{\end{array}}

\newcommand{\be}{\begin{equation}}
\newcommand{\ee}{\end{equation}}

\def\bea{\begin{eqnarray}}
\def\eea{\end{eqnarray}}

\begin{document}

\title{Approximate Nash Equilibria Algorithms for Shapley Network Design Games}

\titlerunning{Approximate Nash Equilibria Algorithms for Shapley Network Design Games}

\author{Hangxin Gan\inst{1}
\and
Xianhao Meng\inst{2}
\and
Chunying Ren\inst{2}\thanks{Corresponding author.\\
\textit{Eamil addresses}: hangxin.gan@mail.nankai.edu.cn (Hangxin Gan),\\ 1120220006@mail.nankai.edu.cn(Xianhao Meng), Rcy9820230019@nankai.edu.cn (Chunying Ren), shi@nankai.edu.cn(Yongtang Shi) }
\and
Yongtang Shi\inst{2}
}
\authorrunning{H. Gan, X. Meng, C. Ren, Y. Shi}

\institute{School of Mathematical Science, Nankai University, Tianjin 300071, P.R. China \and
Center for Combinatorics and LPMC, Nankai University, Tianjin 300071, P.R. China
}
\maketitle

\begin{abstract}

We consider a weighted Shapley network design game, where selfish players choose paths in a network to minimize their cost. The cost function of each edge in the network is affine linear with respect to the sum of weights of the players who choose the edge. We first show the existence of $\alpha$-approximate pure Nash equilibrium by constructing a potential function and establish an upper bound $O(\log_2(W))$ of $\alpha$, where $W$ is the sum of the weight of all players. Furthermore, we assume that the coefficients of the cost function (affine linear function) of the edge all are $\phi$-smooth random variables on $[0,1]$. In this case, we show that $\epsilon$-best response dynamics can compute the $(1+\epsilon)\alpha$-approximate pure Nash equilibrium ($\epsilon$ is a positive constant close to 0) in polynomial time by proving the expected number of iterations is polynomial in $\frac{1}{\epsilon},\phi$, the number of players and the number of edges in the network.

\end{abstract}

\section{Introduction}\label{sec1}

Along with the development of many networks including the Internet, the number of players using networks is increasing, all driven by self-interest and optimize an individual objective function. This naturally leads us to study the behavior of these non-cooperative players in the game, including \cite{ack,rwxy,bcfm,efm}. Specifically, the game is defined on a directed network and each player needs to choose a path from a source vertex to a sink vertex. Each edge in the network has a cost and the players using the edge share the cost equally or according to their weights. All players want to reduce personal cost and choose the path simultaneously, forming a profile of the game. The stable profiles are called \textit{Nash equilibrium}, in which no player has an incentive to unilaterally change strategy. Rosenthal \cite{rwr} first studied the equilibrium of the game on a network and proved the existence of \textit{pure Nash equilibrium} (PNE) in the game. Then Anshelevich et al. \cite{adktwr} named the game \textit{network design with Shapley cost sharing}, also known as \textit{Shapley network design game} or \textit{fair cost sharing game} since it applies the cost sharing mechanism.

Furthermore, the study of PNE gives rise to a number of new issues. Specifically, since the overall performance of the system brought about by the selfish players making their own decisions may be worse than that brought about by centralized authority decision-making, we need an indicator to measure the degree to which the global performance of the system has deteriorated. Papadimitriou \cite{chp} first proposed using the term \textit{price of anarchy} to refer to the deterioration of system performance. Moreover, multiple PNEs may exist in some types of games, but the price of anarchy only studies the deterioration in the worst PNE. Then Anshelevich et al. \cite{adktwr} introduced the term \textit{price of stability} to refer to the ratio of the cost of best Nash equilibrium to the optimum network cost. In \cite{adktwr} they proved that a Shapley network design game with $n$ players always exists a PNE to which the price of stability with respect at most $H(n)=\sum^n_{i=1}\frac{1}{i}$ by using \textit{potential function} introduced by Monderer and Shapley \cite{dmls}.

However, Anshelevich et al. \cite{adktwr} only proved few results when players are \textit{weighted}, including proving the existence of PNE in all weighted Shapley network design game with two players. Chen and Roughgarden \cite{hctr} presented a three-player weighted Shapley network design game that does not admit a PNE, which motivates subsequent study on \textit{$\alpha$-approximate pure Nash equlibrium}  ($\alpha$-APNE). In our model, $\alpha$-APNE is a profile that no players can unilaterally change strategy to reduce the cost by a factor of at least $\alpha$. Then Chen and Roughgarden \cite{hctr} showed that the existence of $O(\log w_{max})$-APNE in every weighted Shapley network design game and proved the price of stability with respect to the $O(\log w_{max})$-APNE is $O(\log W)$, where $w_{max}$ and $W$ are the maximum weight and the sum of weights of all players respectively. In addition, they constructed a network without $o(\frac{\log w_{max}}{\log\log w_{max}})$-APNE.

On the other hand, computing a Nash equilibrium is essential when the game is given from a computational perspective. Apart from proposing Shapley network design game, Anshelevich et al. \cite{adktwr} showed that \textit{best response dynamics} may take exponential time to converge to a PNE. Specifically, best response dynamics is an algorithm to find PNE ($\alpha$-APNE) by picking a player who has a strategy that can reduce personal cost (by more than $\alpha$ times) and updating the strategy as long as the profile is not a PNE ($\alpha$-APNE). Furthermore, Syrgkanis \cite{vs} showed that computing a PNE in the unweighted Shapley network design game is PLS-complete by the reduction from MAX CUT. Therefore, it is necessary to make further assumption on the game in order to ensure an algorithm that can find PNE or $\alpha$-APNE in a polynomial time. Giannakopoulos \cite{yg} assumed the constant cost $c_e$ of the edge as $\phi$-smooth independent random variables on $[0,1]$, which means its probability density function is upper bounded by $\phi$. And Giannakopoulos \cite{yg} proved that best response dynamics can compute the $\alpha$-APNE in polynomial time under this assumption.

In real situations, the cost of each edge in the network is not always constant, but changes with the number of players (or the sum of the weights of players) using the edge. However, to the best of our knowledge, no positive results exist regarding the weighted Shapley network design game where the cost of each edge is not a constant. Therefore, we extend the cost function of each edge to linear function to the total weights of the players that choose the edge. Under this assumption, we prove the existence of $\alpha$-APNE, put forward an algorithm to compute $\alpha$-APNE and analyze its time complexity. 

\subsection{Related Works}


Epstein et al. \cite{efm} proved the existence of \textit{strong Nash equilibria} of unweighted Shapley network design game, where strong Nash equilibria are profiles from which no group of players can decrease the cost of each of its members. Moreover, Epstein et al. \cite{efm} showed that the \textit{strong price of anarchy} of a fair cost sharing game with $n$ players is at most $H(n)=\sum^n_{i=1}\frac{1}{i}$, namely $H(n)$ is the $n$-th harmonic number.

For undirected Shapley network design game with $n$ players, Anshelevich et al. \cite{adktwr} showed that the upper bound of the price of stability is $H(n)=\sum^n_{i=1}\frac{1}{i}$. Furthermore, Christodoulou et al. \cite{cclps} present a lower bound of the price of stability of $42/23\approx1.8261$, and was improved to $348/155\approx2.245$ by Bilò et al. \cite{bcfm}. Fiat et al. \cite{fklos} further restricted the network such that there is a common source vertex and a player in every vertex, and proved the price of stability of the game is $O(\log\log n)$. This further restricted game is called as \textit{broadcast network design game}, and the price of stability of the game was improved to constant by Bilò et al. \cite{vml}.

When the game does not apply the cost sharing mechanism, Fabrikant et al. \cite{ack} considered the game in which each player bears the whole cost of each chosen edge and named it the \textit{network congestion game}. They proved that computing a Nash equilibrium is PLS-complete. Moreover, when all players have the same source vertex and sink vertex, namely the game is \textit{symmetric}, Fabrikant et al. \cite{ack} showed that there is a polynomial algotithm for finding a PNE. Furthermore, Ackermann et al. \cite{arv} showed that the PLS-completeness holds in undirected network congestion game.





\subsection{Our contributions}
Our model is the first to extend the cost function to affine linear function in weighted Shapley network design game proposed by Anshelevich et al. \cite{adktwr}, which is a constant $c_e$ for each edge $e$ in the network. In our model, the cost function of each edge in the network $c_e(w_e)=a_ew_e+b_e$, where $a_e,b_e>0$ are only related to the edge $e$ and $w_e$ is the total weight of the players that choose a path containing edge $e$. By constructing an $\alpha$-approximate potential function $\Phi (P)=\sum_{e\in E}c_e(w_e)\log_2(1+w_e)$, we prove the existence of $\alpha$-APNE, where $\alpha=2\log_2(1+W+w_{max})=O(\log_2(W))$, where $W$ and $w_{max}$ is the sum and maximum of the weight of all players respectively.

To compute the $\alpha$-APNE, we adopt $\epsilon$-best response dynamics which can obtain a $(1+\epsilon)\alpha$-APNE. Under the assumption that $\{a_e,b_e\}_{e\in E}$ are $\phi$-smooth random variables, which means that the probability density function of $a_e,b_e$ is upper bounded by a constant $\phi$, we prove the expected number of iterations is polynomial in $\frac{1}{\epsilon},\phi$, the number of players $n$ and the number of edges $m$ in the network. Specifically, we find out that the expected number of iterations can be bounded by $O(\frac{1}{\epsilon}W^3\phi n^2 m \cdot ln(W \phi nm))$. Since in every step of iteration we can adopt the shortest-path search to verify the existence of $(1+\epsilon)\alpha$-improving deviation, then $\epsilon$-best response dynamics can compute the $(1+\epsilon)\alpha$-APNE in polynomial time. 

Following the introduction, Section \ref{sec2} defines the model and several concepts as preliminaries. Section \ref{sec3} proves the existence of $\alpha$-APNE. Section \ref{sec4} proposes an algorithm to compute the $\alpha$-APNE in polynomial time. Conclusions and possible directions in the future are presented in Section \ref{sec5}. An appendix gathers the mathematical techniques used in the main text. 

\section{Model and Preliminaries}\label{sec2}

We represent a \textit{weighted Shapley network design game} $\Gamma$ by tuple $(\mathcal{G},\mathcal{N},\mathcal{P},\mathcal{C},\mathcal{W})$ with components defined below.

\begin{itemize}

\item \textbf{P1:} $\mathcal{G}=(V,E)$ is a directed graph, where $V$ is the vertex set of $\mathcal{G}$ and $E$ is the edge set of $\mathcal{G}$. Here we further define $m=|E|$, which is the number of edges in graph $\mathcal{G}$.

\item \textbf{P2:} $\mathcal{N}$ is a collection of $n$ different players $(n\geq2)$. Moreover, each player $i \in \mathcal{N}$ is identified with a source-sink vertex pair $(s_i,t_i)$.

\item \textbf{P3:} $\mathcal{P}=(\mathcal{P}_i)_{i \in \mathcal{N}}$, where $\mathcal{P}_i$ denote the set of simple $s_i-t_i$ paths. Furthermore, each player $i$ chooses a path $P_i \in \mathcal{P}_i$ from its source vertex to its sink vertex.

\item \textbf{P4:} $\mathcal{C}=(c_e)_{e \in E}$, where $c_e(\cdot)$ is the non-negative cost function of the edge $e$.

\item \textbf{P5:} $\mathcal{W}=(w_i)_{i \in \mathcal{N}}$, where each component $w_i>0$ is a positive weight of player $i \in \mathcal{N}$. When $w_i=w_j$ for any $i,j \in \mathcal{N}$, then $\Gamma$ is called unweighted. Otherwise, $\Gamma$ is called weighted.
\end{itemize}

All players are selfish, and each of them chooses a path independently. Then it results in a profile $P=(P_i)_{i \in \mathcal{N}}$, where $P_i$ denotes the simple $s_i-t_i$ path used by player $i \in \mathcal{N}$. For a given profile and an edge $e \in E$, we define $U_e(P):=\{i|i \in \mathcal{N}, e \in P_i\}$ as the ``user'' of edge $e$. Moreover, we define $w_e=\sum_{i \in U_e(P)} w_i$ as the total weight of the players that choose a path containing edge $e$. Furthermore, for a given profile and a player $i$, the cost share $c^i_e$ of an edge $e \in E$ is $c_e(\cdot)\cdot w_i/w_e$. Then the cost to player $i$ in a profile is the sum of its cost shares: $c_i(P)=\sum_{e\in P_i} c^i_e$. In addition, the cost $c(P)$ of a profile $P$ is defined as $\sum_{i \in \mathcal{N}} c_i(P)$, which can be also written as $\sum_{e \in \cup_{i \in \mathcal{N}}P_i}c_e(\cdot)$. 

To state our discussion precisely, for each player $i \in \mathcal{N}$, we denote by $P_{-i}$ a sub-profile $(P_{i'})_{i'\in \mathcal{N}\backslash\{i\}}$ of strategies used by ``opponents" $i'\in \mathcal{N}\backslash\{i\}$ of player $i$, and use $P=(P_i,P_{-i})$ when we need to show explicitly the strategy $P_i$ used by player $i$ . 

\begin{definition}\label{def1} \textbf{(pure Nash equilibrium)}
Given an instance $(\mathcal{G},\mathcal{N},\mathcal{P},\mathcal{C},\mathcal{W})$ of the game $\Gamma$. We call a profile $P=(P_i)_{i \in \mathcal{N}}$ a pure Nash equilibrium(PNE) if, for each $i$, $P_i$ minimizes $c_i(P)$ over all paths in $\mathcal{P}_i$ while keeping $P_j$ fixed for $j\neq i$. That is:
\begin{align}\label{eqn1}
c_i(P)=c_i(P_i,P_{-i})\leq c_i(P'_i,P_{-i})  
\end{align}
for any $i \in \mathcal{N}$, any $P'_i \in \mathcal{P}_i$.
\end{definition}

In fact, Inequality (\ref{eqn1}) states that $P_i$ is a \textit{best-response} to $P_{-i}$ for player $i$. Therefore, there is no incentive for any player $i \in \mathcal{N}$ to unilaterally change strategy when the profile is already a PNE.

To prove PNE exists, a common approach is to prove that there is a \textit{potential function} in the game. Here we give the definition of the potential function.

\begin{definition}\label{def3} \textbf{(exact potential function)}
Given an instance $(\mathcal{G},\mathcal{N},\mathcal{P},\mathcal{C},\mathcal{W})$ of the game $\Gamma$. We call a real-valued function $\Phi:\prod_{i\in\mathcal{N}}\mathcal{P}_i\rightarrow[0,\infty)$ an (exact) potential function if
\begin{align}
    \Phi(P'_i,P_{-i})-\Phi(P_i,P_{-i})=c_i(P'_i,P_{-i})-c_i(P_i,P_{-i})
\end{align}
for any $i\in \mathcal{N}$, any $P_i,P'_i\in\mathcal{P}_i$, any $P_{-i}\in\prod_{j\in\mathcal{N}\backslash\{i\}}\mathcal{P}_j$.
\end{definition}

If there is a potential function in a finite game, since the potential function quantifies the cost reduction of a unilateral change of strategy exactly, then the global minimums of the potential function are PNEs of the game.

However, in some cases, the game may not admit a PNE. Then we have to consider a relaxed notion of \textit{$\alpha$-approximate pure Nash equilibrium} instead.

\begin{definition}\label{def2} \textbf{($\alpha$-approximate pure Nash equilibrium) }
Given an arbitrary instance $(\mathcal{G},\mathcal{N},\mathcal{P},\mathcal{C},\mathcal{W})$ of the game $\Gamma$ and an arbitrary constant $\alpha \geq 1$.  We call a profile $P=(P_i)_{i \in \mathcal{N}}$ an $\alpha$-approximate pure Nash equilibrium ($\alpha$-APNE) if
\begin{align}\label{eqn2}
c_i(P)=c_i(P_i,P_{-i})\leq \alpha \cdot c_i(P'_i,P_{-i})  
\end{align}
for any $i \in \mathcal{N}$, any $P'_i \in \mathcal{P}_i$.
\end{definition}

In fact, Inequality (\ref{eqn2}) states that a unilateral change of strategy reduces the cost at a rate of at most $\frac{c_i(P_i,P_{-i})-c_i(P'_i,P_{-i})}{c_i(P_i,P_{-i})} \leq 1-\frac{1}{\alpha}$ in an $\alpha$-APNE. Moreover, we call a deviation \textit{$\alpha$-improving} if the deviation decreases the cost incurred by the player by at least an $\alpha$ multiplicative factor. Therefore, $\alpha$-APNE is a profile from which no $\alpha$-improving deviations exist. We further point out that the smaller the constant $\alpha$ is, the closer the profile $P$ would approximate a PNE.

Correspondingly, we can prove the existence of $\alpha$-APNE by finding the approximate potential function for the game.

\begin{definition}\label{def4} \textbf{($\alpha$-approximate potential function)}
Given an instance $(\mathcal{G},\mathcal{N},\mathcal{P},\mathcal{C},\mathcal{W})$ of the game $\Gamma$. For any player $i\in \mathcal{N}$, $P_i,P'_i\in\mathcal{P}_i$ satisfies $P_i\rightarrow P'_i$ is an $\alpha$-improving deviation. We call a real-valued function $\Phi:\prod_{i\in\mathcal{N}}\mathcal{P}_i\rightarrow[0,\infty)$ an $\alpha$-approximate potential function if
\begin{align}
    \Phi(P'_i,P_{-i})-\Phi(P_i,P_{-i})\leq \alpha c_i(P'_i,P_{-i})-c_i(P_i,P_{-i})
\end{align}
for any $P_{-i}\in\prod_{j\in\mathcal{N}\backslash\{i\}}\mathcal{P}_j$.
\end{definition}

If there is a $\alpha$-approximate potential function in a finite game, since the potential function strictly decreases in every $\alpha$-improving deviation, then there will be no $\alpha$-improving deviation after a finite number of steps, which means the game admits an $\alpha$-APNE.

Without loss of generality, we assume $\min_{i \in \mathcal{N}} w_i=1$ in the following discussion, and we define $w_{max}=\max_{i \in \mathcal{N}} w_i$ and $W=\sum_{i \in \mathcal{N}} w_i$ to simplify our discussion.

\section{Existence of $\alpha$-APNE}\label{sec3}

Anshelevich et al. \cite{adktwr} proved that the unweighted Shapley network design game admits a PNE, and pointed out that weighted Shapley network design game does not admit a PNE. Since our model is an extension of weighted Shapley network design game, we have to turn our attention to $\alpha$-APNE instead of PNE. 

\begin{theorem} \label{exist}
Given an instance $(G,\mathcal{N},\mathcal{P},\mathcal{C},\mathcal{W})$ of the game $\Gamma$. If the cost function is affine linear, namely $c_e(w_e)=a_ew_e+b_e$, where $a_e,b_e>0$ are only related to the edge $e$ and $w_e$ is the total weight of the players that choose a path containing edge $e$, then the game admits an $\alpha$-APNE, where $\alpha=O(\log_2(W))$.
\end{theorem}

\begin{proof}
    For a weighted Shapley network design game $\Gamma$, we define a function $\Phi$ as follows:
    \begin{align*}
        \Phi (P)=\sum_{e\in E}c_e(w_e)\log_2(1+w_e).     
    \end{align*}
     Now we consider an $\alpha$-improving deviation of player $i$ from the profile $P=(P_i,P_{-i})$ to the profile $P'=(P'_i,P_{-i})$. Without loss of generality, we can assume that $P_i$ and $P'_i$ are disjoint. Otherwise, we can replace $P_i$ and $P'_i$ with $P_i\backslash P'_i$ and $P'_i\backslash P_i$ in the following discussion. By the definition of $\alpha$-improving, we have
    \begin{align*}
        \alpha \cdot \sum_{e\in P'_i}{c_e(w_e+w_i)} \cdot \frac{w_i}{w_e+w_i} \leq \sum_{e\in P_i}{c_e(w_e)} \cdot \frac{w_i}{w_e}.
    \end{align*}
    Here, $w_e$ denotes the total weight on edge $e$ before the deviation and we let $\alpha=2\log_2(1+W+w_{max})=O(\log_2(W))$. Since $c_e(w_e)=a_ew_e+b_e$, we have
    \begin{align} \label{eqn3}
         \alpha \cdot \sum_{e\in P'_i}{[a_e(w_e+w_i)+b_e]} \cdot \frac{w_i}{w_e+w_i} \leq \sum_{e\in P_i}{(a_ew_e+b_e)} \cdot \frac{w_i}{w_e}.
    \end{align}
    
    Then we can derive the following:
    \begin{align*}
        \Delta \Phi&= \sum_{e\in P'_i}{(a_ew_e+b_e)}[\log_2(1+w_e+w_i)-\log_2(1+w_e)]
        \\&\,\quad -\sum_{e\in P_i}{(a_ew_e+b_e)}[\log_2(1+w_e)-\log_2(1+w_e-w_i)]
        \\&\,\quad +\sum_{e\in P'_i}a_ew_i\log_2(1+w_e+w_i)-\sum_{e\in P_i}a_ew_i\log_2(1+w_e-w_i) 
        \\&<\sum_{e\in P'_i}{(a_ew_e+b_e)}\log_2[e(1+w_i)]\frac{w_i}{w_e+w_i}\nonumber
        \\&\,\quad -\sum_{e\in P_i}{(a_ew_e+b_e)}\frac{w_i}{w_e}+\sum_{e\in P'_i}a_ew_i\log_2(1+w_e+w_i)
        \\&=\sum_{e\in P'_i}a_ew_i\{\log_2(1+w_e+w_i)+\log_2[e(1+w_i)]\frac{w_e}{w_e+w_i}\}
        \\&\,\quad +\sum_{e\in P'_i}b_e\log_2[e(1+w_i)]\frac{w_i}{w_e+w_i}-\sum_{e\in P_i}{(a_ew_e+b_e)}\frac{w_i}{w_e}
        \\&<\alpha \sum_{e\in P'_i}{[a_e(w_e+w_i)+b_e]}\cdot \frac{w_i}{w_e+w_i}-\sum_{e\in P_i}{(a_ew_e+b_e)} \cdot \frac{w_i}{w_e}
        \\&=\alpha c_i(P')-c_i(P).    
    \end{align*}
    For the first inequality, we take the logarithm of $e(1+w_i)\geq(1+\frac{w_i}{w_e+1})^\frac{w_e+1}{w_i}(1+\frac{w_i}{w_e+1})\geq(1+\frac{w_i}{w_e+1})^\frac{w_e+w_i}{w_i}$ and apply $\log_2(\frac{1+w_e}{1+w_e-w_i})>1\geq\frac{w_i}{w_e}$ when $2w_i\geq1+w_e$ or take the logarithm of $(1+\frac{w_i}{1+w_e-w_i})^\frac{w_e}{w_i}\geq(1+\frac{w_i}{1+w_e-w_i})^\frac{1+w_e-w_i}{w_i}\geq2$ when $2w_i<1+w_e$. Moreover, we use $\alpha\geq\max\{\log_2[e(1+w_i)],\log_2(1+w_e+w_i)+\log_2[e(1+w_i)]\frac{w_e}{w_e+w_i}\}$ for the last inequality. Therefore, we have proved
    \begin{align} \label{eqn4}
        \Phi(P')-\Phi(P)<\alpha c_i(P')-c_i(P),
    \end{align}
    which suggests that $\alpha$-improving deviations strictly decrease the function $\Phi$. Since the number of different profiles are finite, the theorem can be directly derived.
\end{proof}

\section{Compute the $(1+\epsilon)\alpha$-APNE}\label{sec4}


    Initially we propose an algorithm to find the $(1+\epsilon)\alpha$-APNE of the game $\Gamma$ as followings, where $\epsilon$ is a positive constant close to 0.

    \begin{algorithm}[h] 
	\renewcommand{\algorithmicrequire}{\textbf{Input:}}
    \renewcommand{\algorithmicensure}{\textbf{Output:}}
	\caption{$\epsilon$-approximate best response dynamics($\epsilon$-ABRD)} \label{alg2}
    \begin{description}
        \item[Input:] A weighted Shapley network design game $\Gamma=(\mathcal{G},\mathcal{N},\mathcal{P},\mathcal{C},\mathcal{W})$; a strategy profile $P \in \mathcal{P}$
        \item[Output:] A $(1+\epsilon)\alpha$-APNE of game $\Gamma$ 
    \end{description}
    \begin{enumerate}
    \item \textbf{While $P$} is not an $(1+\epsilon)\alpha$-APNE \textbf{do}
    \item \quad Choose $i \in \mathcal{N}$, $P'_i \in \mathcal{P}_i$ such that $(1+\epsilon)\alpha c_i(P'_i,P_{-i})<c_i(P)=c_i(P_i,P_{-i})$ 
    \item \quad $P\leftarrow (P'_i,P_{-i})$
    \item \textbf{end while}
    \item \textbf{return} $P$
\end{enumerate}
\end{algorithm}

Specifically, the first and the second line of the Algorithm \ref{alg2} is finding an $(1+\epsilon)\alpha$-improving deviation for player $i$. For a given instance of game $\Gamma$, we may not be able to search exhaustively in the strategy space $\mathcal{P}_i$ since it will cost exponential time. Instead, we can make a shortest-path search for player $i$, which can find the best response for player $i$ efficiently and cost polynomial time. In this way, we only need to verify whether the best response is an $(1+\epsilon)\alpha$-improving deviation, which can also be computed in polynomial time. Therefore, in order to show the Algorithm \ref{alg2} runs in polynomial time, we only need to prove the number of iterations is polynomial in the given parameters. Before we begin to prove, we first make an assumption of the parameter and propose a lemma as preparation.

\begin{definition}
    Let $X$ be a random variable, we say $X$ is a $\phi$-smooth random variable if and only if $X$ is defined on $[0,1]$ and the probability density function of $X$ is up to $\phi$($\phi \geq 1$).    
\end{definition}

    In our model, it is natural to assume $\{a_e,b_e\}_{e\in E}$ are $\phi$-smooth random variables. For $\phi$-smooth random variable, we can derive the lemma below. 
    
\begin{lemma} \label{lem2}
    Let $X_1,X_2,...,X_n;Y_1,Y_2,...,Y_n(n\geq2)$ be $\phi$-smooth independent random variables on [0,1]. Let $g(x)=\min\{\frac{\alpha}{x}ln\frac{\alpha}{x},n^\lambda\}$. Then for any $\lambda \in N$ and $\alpha\geq1$, we have
    \begin{align} \label{eqn66}
        \mathbb{E}[g(\min_{i\in[n]}\{X_i+Y_i\})] \leq\alpha\phi(n^2+1) \cdot ln(\alpha\phi n)
    \end{align}
\end{lemma}

\begin{proof}
    For two random variables $S,T$, we define $S\preceq T$ if and only if $F_T(x)\leq F_S(x)$ for any $x \in \mathbb{R}$, where $F_S(x),F_T(x)$ are the \textit{cumulative distribution function} of $S,T$ respectively. To prove the lemma, we define two sets of independent uniformly distributed random variables $S_1,S_2,...,S_n;T_1,T_2,...,T_n$ on $[0,\frac{1}{\phi}]$. In addition, we denote $[n]=\{1,2,...n\}$ to simplify our discussion.
\begin{claim}
    For any $i\in[n]$, we have 
    \begin{align*}
        min_{i\in[n]}S_i \preceq min_{i\in[n]}X_i, min_{i\in[n]}T_i \preceq min_{i\in[n]}Y_i.
    \end{align*}
\end{claim}

    To prove the claim, we firstly prove $F_{X_i}(t)\leq F_{S_i}(t)$ for any $t \in \mathbb{R}$, then we can derive $F_{Y_i}(t)\leq F_{T_i}(t)$ similarly.

    Since $S_i$ is defined on $[0,\frac{1}{\phi}]$, we have $F_{S_i}(t)=1$ for any $t\geq\frac{1}{\phi}$, then we have $F_{X_i}(t)\leq F_{S_i}(t)$ definitely. When $0\leq t\leq\frac{1}{\phi}$, since $X_i$ is a $\phi$-smooth random variable for any $i\in[n]$, its probability density function is upper-bounded by $\phi$. Then we have
    \begin{align*}
        F_{X_i}(t)\leq\int^t_0 \phi dx= F_{S_i}(t)
    \end{align*}
    for all $0\leq t\leq\frac{1}{\phi}$. Therefore, we have $F_{X_i}(t)\leq F_{S_i}(t)$ and $F_{Y_i}(t)\leq F_{T_i}(t)$. Furthermore, since $X_1, X_2,..., X_n$ are independent random variables, we have
    \begin{align*}
    F_{min_{i\in[n]}X_i}(t)&=1-\prod^n_{i=1}(1-F_{X_i}(t))\leq1-\prod^n_{i=1}(1-F_{S_i}(t))=F_{min_{i\in[n]}S_i}(t),
    \end{align*}
    which completes the proof of the claim.

    Moreover, we can find that $g(x)=\min\{\frac{\alpha}{x}ln\frac{\alpha}{x},n^\lambda\}$ is nonincreasing with respect to $x$. From the Proposition 9.1.2 in \cite{js}, we have 
    \begin{align} \label{eqn10} 
        \mathbb{E}[g(\min_{i\in[n]}\{X_i+Y_i\})]\leq\mathbb{E}[g(\min_{i\in[n]}\{S_i+T_i\})].
    \end{align}
    We define $W=\frac{1}{\min_{i\in[n]}\{S_i+T_i\}}$, then $W\in[\frac{\phi}{2},\infty)$. Since $S_i,T_i$ are independent uniformly distributed random variables on $[0,\frac{1}{\phi}]$, we have
    \begin{eqnarray}
         F_W(w)=\prod^n_{i=1} Prob[S_i+T_i\geq\frac{1}{w}]
            =\left\{
            \begin{aligned}
            &(1-\frac{\phi^2}{2w^2})^n, \quad\quad\quad\ \ w\in[\phi,\infty)\\
            &(2-\frac{2\phi}{w}+\frac{\phi^2}{2w^2})^n, \quad w\in[\frac{\phi}{2},\phi)
            \end{aligned}.
            \right.
    \end{eqnarray}
    The last equation is due to Lemma \ref{lem6} where we let $A \leftarrow \frac{1}{\phi}, t \leftarrow \frac{1}{w}$.
    
    For $i=1,2...,\lambda,\lambda+1$, we define $Q_i$ as the random event that $W\in[\phi n^{i-1},\phi n^{i})$. Moreover, we let $Q_0$ and $Q_{\lambda+2}$ denote the random event that $W\in[\frac{\phi}{2},\phi)$ and $W\in[\phi n^{\lambda+1}, \infty)$ respectively. Then, we can derive the following upper bound on the expectation in Inequality (\ref{eqn10}) by:
    \begin{align*}
         \mathbb{E}[g(\min_{i\in[n]}\{X_i+Y_i\})]&\leq
         \mathbb{E}[\min\{\alpha\cdot Wln(\alpha\cdot W),n^\lambda\}]\\
         &\leq\sum^{\lambda+1}_{i=0}Prob[Q_i]\cdot\mathbb{E}[\alpha\cdot Wln(\alpha\cdot W)|Q_i]+Prob[Q_{\lambda+2}]\cdot n^\lambda\\
         &\leq\sum^{\lambda+1}_{i=1}[F_W(\phi n^{i})-F_W(\phi n^{i-1})]\cdot  \alpha\phi n^{i} \cdot ln(\alpha\phi n^{i})\\
         &\,\quad +[F_W(\phi)-F_W(\frac{\phi}{2})]\alpha\phi ln(\alpha\phi) +[1-F_W(\phi n^{\lambda+1})]n^\lambda\\
         &\leq\alpha\phi\cdot ln(\alpha\phi n)\cdot\sum^{\lambda+1}_{i=1}[F_W(\phi n^{i})-F_W(\phi n^{i-1})]\cdot  n^{i} \cdot i \\
         &\,\quad +[F_W(\phi)-F_W(\frac{\phi}{2})]\alpha\phi ln(\alpha\phi) +[1-F_W(\phi n^{\lambda+1})] n^\lambda\\
         &=\alpha\phi\cdot ln(\alpha\phi n)\cdot\sum^{\lambda+1}_{i=1}[(1-\frac{1}{2n^{2i}})^n-(1-\frac{1}{2n^{2i-2}})^n]\cdot in^{i} \\
         &\,\quad +(\frac{1}{2})^n\cdot\alpha\phi\cdot ln(\alpha\phi) + [1-(1-\frac{1}{2n^{2\lambda+2}})^n]\cdot n^\lambda \\
         &\leq\alpha\phi\cdot ln(\alpha\phi n)\cdot\sum^{\lambda+1}_{i=1}[\frac{1}{1+\frac{n}{2n^{2i}}}-(1-\frac{n}{2n^{2i-2}})]\cdot  in^{i}\\
         &\,\quad +(\frac{1}{2})^n\cdot\alpha\phi\cdot ln(\alpha\phi) + [1-(1-\frac{n}{2n^{2\lambda+2}})]\cdot n^\lambda \\
         &\leq\alpha\phi\cdot ln(\alpha\phi n)\cdot\sum^{\lambda+1}_{i=1}\frac{i}{2n^{i-3}} +(\frac{1}{2})^n\cdot\alpha\phi\cdot ln(\alpha\phi) + \frac{1}{2n^{\lambda+1}} \\
         &\leq\alpha\phi \cdot ln(\alpha\phi n)\cdot (n^2+1).
    \end{align*}
    The second inequality is due to the law of total expectation. And the third to last inequality is due to Lemma \ref{lem5}.
    Therefore, we complete the proof of the lemma.
\end{proof}

    It is worth mentioning that there is no $\lambda$ on the right side of Inequality \ref{eqn66}. With the formula for changing base, the second term of the function $g$ has almost no effect on the conclusion.
    
    Now we begin to show that $\epsilon$-ABRD always find the $(1+\epsilon)\alpha$-APNE in a polynomial time by proving the expected number of iterations is polynomial in the given parameters. 
    
\begin{theorem}
    Given an instance $(G,\mathcal{N},\mathcal{P},\mathcal{C},\mathcal{W})$ of the game and $c_e(w_e)=a_ew_e+b_e$, where $\{a_e,b_e\}_{e\in E}$ are $\phi$-smooth random variables. Then, the expected number of iterations of $\epsilon$-ABRD is at most $O(\frac{1}{\epsilon}W^3\phi n^2 m \cdot ln(W \phi nm))$.
\end{theorem}

\begin{proof}
    We recall that the approximate potential function $\Phi$ of the game $\Gamma$ is
    \begin{align*}
        \Phi (P)=\sum_{e\in E}c_e(w_e)\log_2(1+w_e),     
    \end{align*}
    which can be upper-bounded by
    \begin{align} \label{eqn5}
        \Phi (P) \leq \sum_{e\in E}c_e(W)\log_2(1+w_e)\leq m\max_{e\in E}\{a_eW+b_e\}\log_2(1+W).
    \end{align}
    To simplify our discussion, we let $c_{max}=\max_{e\in E}\{a_eW+b_e\}$. Since $\min_{i\in \mathcal{N}}w_i=1$ and each player uses at least one edge, then $\Phi$ can be lower bounded by
    \begin{align} \label{eqn6}
        \Phi (P) \geq \sum_{e\in E} \min_{e\in E}\{a_e+b_e\}\log_2(1+w_e) \geq \min_{e\in E}\{a_e+b_e\}\log_2(1+W).
    \end{align}
    The second inequality is due to $\log_2(1+x)+\log_2(1+y)\geq \log_2(1+x+y)$ for any $x,y>0$.
    To simplify our discussion, we let $c_{min}=\min_{e\in E}\{a_e+b_e\}$. 
    
    Furthermore, according to algorithm \ref{alg2}, we assume that we move $T$ steps $P^0 \rightarrow P^1 \rightarrow ... \rightarrow P^T$ during the execution of our $\epsilon$-ABRD and $P \rightarrow P'$ is one of them. Since $c_i(P)\geq \frac{c_{min}}{W}$, we can derive the following from Inequality (\ref{eqn4}) and Inequality (\ref{eqn5}):
    \begin{align}
        \Phi(P)-\Phi(P')&>c_i(P)-\alpha c_i(P')\nonumber\\&>\frac{\epsilon}{1+\epsilon}c_i(P)\nonumber
        \\&\geq \frac{\epsilon}{(1+\epsilon)W}c_{min}\nonumber
        \\&\geq \frac{\epsilon c_{min}}{(1+\epsilon)\cdot W\cdot \log_2(1+W)\cdot m\cdot c_{max}}\Phi(P).
    \end{align}
    Now we have
    \begin{align} \label{eqn7}
        \Phi(P')<(1-\epsilon')\Phi(P),\quad \text{where}\; \epsilon'=\frac{\epsilon c_{min}}{(1+\epsilon)W\log_2(1+W)mc_{max}}.
    \end{align}
    Since Inequality (\ref{eqn7}) holds for any move during the execution of the $\epsilon$-ABRD, we can derive the following from Inequality (\ref{eqn5}) and Inequality (\ref{eqn6}):
    \begin{align*}
        c_{min}\log_2(1+W)\leq \Phi(P^T)< (1-\epsilon')^T\Phi(P^0)\leq (1-\epsilon')^Tmc_{max}\log_2(1+W).
    \end{align*}
    Therefore, we take logarithms at both sides of the inequality and get 
    \begin{align} \label{eqn8}
        Tln(1-\epsilon')>ln\frac{c_{min}}{mc_{max}}.
    \end{align}
    Since $0<\epsilon'<1$, we can derive the following upper bound on the number of steps for our $\epsilon$-ABRD from Inequality (\ref{eqn8}):
    \begin{align} \label{eqn9}
        T&<-\frac{1}{ln(1-\epsilon')}ln\frac{mc_{max}}{c_{min}}\nonumber\\&\leq\frac{1}{\epsilon'}ln\frac{mc_{max}}{c_{min}}\nonumber\\&=(1+\frac{1}{\epsilon})W\log_2(1+W)\frac{mc_{max}}{c_{min}}ln\frac{mc_{max}}{c_{min}}.
    \end{align}
    The second inequality is due to $\epsilon'\geq ln(\frac{1}{1-\epsilon'})$ when $0<\epsilon'<1$.

    On the other hand, we recall that the approximate potential function $\Phi$ is strictly decreasing at every step of the $\epsilon$-ABRD, which means no profile appears more than once during the $\epsilon$-ABRD. Therefore, the total number of iterations cannot be larger than the number of different profiles, which is up to $(2^n)^m=2^{nm}$. Combining this with Inequality (\ref{eqn9}) we have
    \begin{align}
        T \leq \min\{(1+\frac{1}{\epsilon})W\log_2(1+W)\frac{mc_{max}}{c_{min}}ln\frac{mc_{max}}{c_{min}},2^{nm}\}.
    \end{align}
    Since $c_{min}=\min_{e\in E}\{a_e+b_e\}$ and $\{a_e,b_e\}_{e\in E}$ are $\phi$-smooth random variables on $[0,1]$, then we can derive the following from Lemma \ref{lem2}:
    \begin{align*}
        \mathbb{E}[T]&\leq (1+\frac{1}{\epsilon})W\log_2(1+W)\mathbb{E}[\min\{\frac{m(1+W)}{c_{min}}ln\frac{m(1+W)}{c_{min}},m^{nm}\}]\\
        &\leq (1+\frac{1}{\epsilon})W\log_2(1+W)\cdot m (1+W)\phi(n^2+1) \cdot ln(m(1+W)\phi\cdot n)\\
        &\leq 4\cdot(1+\frac{1}{\epsilon})W^3\cdot\phi\cdot n^2 \cdot m \cdot ln(W\cdot \phi \cdot n \cdot m)\\
        &=O(\frac{1}{\epsilon}W^3\cdot\phi\cdot n^2 \cdot m \cdot ln(W\cdot \phi \cdot n \cdot m)),
    \end{align*}
    which completes the proof of the theorem.
\end{proof}

%

\section{Conclusion}\label{sec5}

Our paper give the first result on weighted Shapley network design game with affine linear cost function $c_e(w_e)=a_ew_e+b_e$. The first major result is the existence of $O(\log_2(W))$-APNE, which is shown by constructing an $O(\log_2(W))$-approximate potential function $\Phi (P)=\sum_{e\in E}c_e(w_e)\log_2(1+w_e)$. For the weighted Shapley network design game with constant cost function $c_e$, Chen and Roughgarden \cite{hctr} proved the existence of $\log_2[e(1+w_{max})]$-APNE. It can be seen from the comparison that the value of $\alpha$ becomes a little larger but is still within the acceptable range.

On the other hand, computing $(1+\epsilon)\alpha$-APNE in polynomial time through $\epsilon$-best response dynamics is another major result. Since we can adopt the shortest-path search to ensure every step of iteration runs in polynomial time, we only need to prove the expected number of iterations is polynomial in parameters. Specifically, We assume $\{a_e,b_e\}_{e\in E}$ are $\phi$-smooth random variables and prove the expected number of iterations is at most $O(\frac{1}{\epsilon}W^3\phi n^2 m \cdot ln(W \phi nm))$, which completes the analysis on the time complexity of $\epsilon$-best response dynamics. For the unweighted Shapley network design game with $\phi$-smooth random variables $c_e$, Giannakopoulos \cite{yg} proved the expected number of iterations of $\epsilon$-best response dynamics is $O(\frac{1}{\epsilon}\phi \log(\phi)n\log^3(n)m^5)$. Then we can find out that the number of $m$ in the expected number of iterations is greatly reduced in our model.

In the future, there are plenty of issues with the model that are worth studying, such as the price of anarchy and the price of stability of the game. Furthermore, we can further generalize the cost function into a polynomial function or even a general continuous function and study the $\alpha$-APNE of the game.

\section*{Acknowledgements}

This work was partially supported by the National Natural Science Foundation of China (No. 12161141006), the Natural Science Foundation of Tianjin (No. 20JCJQJC00090).

\newpage
\section*{Appendix}

\begin{lemma} \label{lem5}
    Let $x\in(0,1),n\in N$, then we have:
        \begin{enumerate}
            \item $(1-x)^n\geq1-nx$;
            \item $(1-x)^n\leq\frac{1}{1+nx}$.
        \end{enumerate}
\end{lemma}

\begin{proof}
    We use mathematical induction to prove part 1. When $n=1$, the left and right sides of the inequality are equal. If the inequality holds when $n=k$, then for $n=k+1$, we have
    \begin{align*}
        (1-x)^{k+1}&=(1-x)^k(1-x)\\
        &>(1-kx)(1-x)\\
        &=1-(k+1)x+kx^2\\
        &>1-(k+1)x.
    \end{align*}
    Therefore, by mathematical induction we prove part 1.

    For part 2, from binomial expansion we have $(1+x)^n\geq1+nx$. Then we can derive the following:
    \begin{align*}
        1>(1-x^2)^n=(1+x)^n(1-x)^n\geq(1+nx)(1-x)^n
    \end{align*}
    Therefore, we prove $(1-x)^n\leq\frac{1}{1+nx}$ and complete the proof of lemma.
\end{proof}

\begin{lemma} \label{lem6}
    Let $X,Y$ be two independent uniformly distributed random variables on $[0,A]$, where $A>0$ is a constant. For a random variable $X$, $F_X(t)$ denotes the cumulative function of $X$. Then we have:
    \begin{eqnarray}
        F_{X+Y}(t)
          =\left\{
            \begin{aligned}
            &\frac{t^2}{2A^2}, \qquad\qquad\quad t\in[0,A)\\
            &\frac{2t}{A}-\frac{t^2}{2A^2}-1, \quad t\in[A,2A]\nonumber
            \end{aligned}.
            \right.
    \end{eqnarray}
    
\end{lemma}

\begin{proof}
    We assume $Z=X+Y$, then $Z$ is defined on $[0,2A]$. For a random variable $X$, let $f_X(t)$ be the probability density function of $X$. Then we have $f_X(t)=f_Y(t)=\frac{1}{A}$ for any $t\in [0,A]$. Since $X,Y$ are independent random variables, we have
    \begin{align} \label{eqn99}
        f_Z(t)=\int^\infty_0f_X(x)f_Y(t-x)dx.
    \end{align}
    When $t\in[0,A)$, from Equation \ref{eqn99} we have
    \begin{align*}
        f_Z(t)=\int^t_0\frac{1}{A^2}dx=\frac{t}{A^2}.
    \end{align*}
    Similarly, we can derive the following from Equation \ref{eqn99} when $t\in[A,2A]$:
    \begin{align*}
        f_Z(t)=\int^A_{t-A}\frac{1}{A^2}dx=\frac{2A-t}{A^2}.
    \end{align*}

    Furthermore, we have
    \begin{eqnarray}
        F_Z(t)=\int^t_0f_Z(x)dx=\left\{
            \begin{aligned}
            &\frac{t^2}{2A^2}, \qquad\qquad\quad t\in[0,A)\\
            &\frac{2t}{A}-\frac{t^2}{2A^2}-1, \quad t\in[A,2A]\nonumber
            \end{aligned}.
            \right.
    \end{eqnarray}
\end{proof}

\end{document}